\documentclass[11pt]{article}
\usepackage{moriond,epsfig}
\usepackage{subfigure}

\def\Journal#1#2#3#4{{#1} {\bf #2}, #3 (#4)}

\def\NIMA{{\em Nucl. Instrum. Methods} A}

\def\PRL{\em Phys. Rev. Lett.}
\def\PRD{{\em Phys. Rev.} D}

\def\be{\begin{equation}}
\def\ee{\end{equation}}
\def\bea{\begin{eqnarray}}
\def\eea{\end{eqnarray}}

\def\fheighta{5.9cm}
\def\fheightb{6.0cm}
\begin{document}
\vspace*{4cm}
\title{Higgs Boson Searches Beyond the Standard Model with ATLAS}

\author{ J.~Kroseberg (on behalf of the ATLAS Collaboration)}

\address{Physikalisches Institut der Universit\"at Bonn, Nu\ss allee 12, D-53115 Bonn, Germany}

\maketitle\abstracts{
Recent searches for Higgs bosons in the context of extensions to the Standard Model of Particle Physics with the ATLAS detector at the Large Hadron Collider are discussed. All presented analyses use data recorded at  a $pp$ center-of-mass energy of 7~TeV in 2011 with integrated luminosities  between 1 and 5~fb$^{-1}$. No significant deviations from the background expectations are found and corresponding constraints on physics beyond the Standard Model are obtained. 
}

\begin{boldmath}
\section{Introduction}
\end{boldmath}
\noindent
While much of the current interest in Higgs boson searches is focussed on the Standard Model (SM) case~\cite{smmoriond},  a number of important Higgs sector scenarios beyond the SM are also being investigated. In the following, selected searches with the ATLAS experiment at the Large Hadron Collider for beyond-SM neutral, charged and doubly-charged Higgs boson are discussed. The analyses are based on data recorded in 2011 at a $pp$ center-of-mass energy of 7 TeV. Signal expectations are derived using information compiled by the LHC Higgs Cross Section WG~\cite{lhcxs}. \\

\begin{figure}[t!]
\hfill
\subfigure[]{\psfig{figure=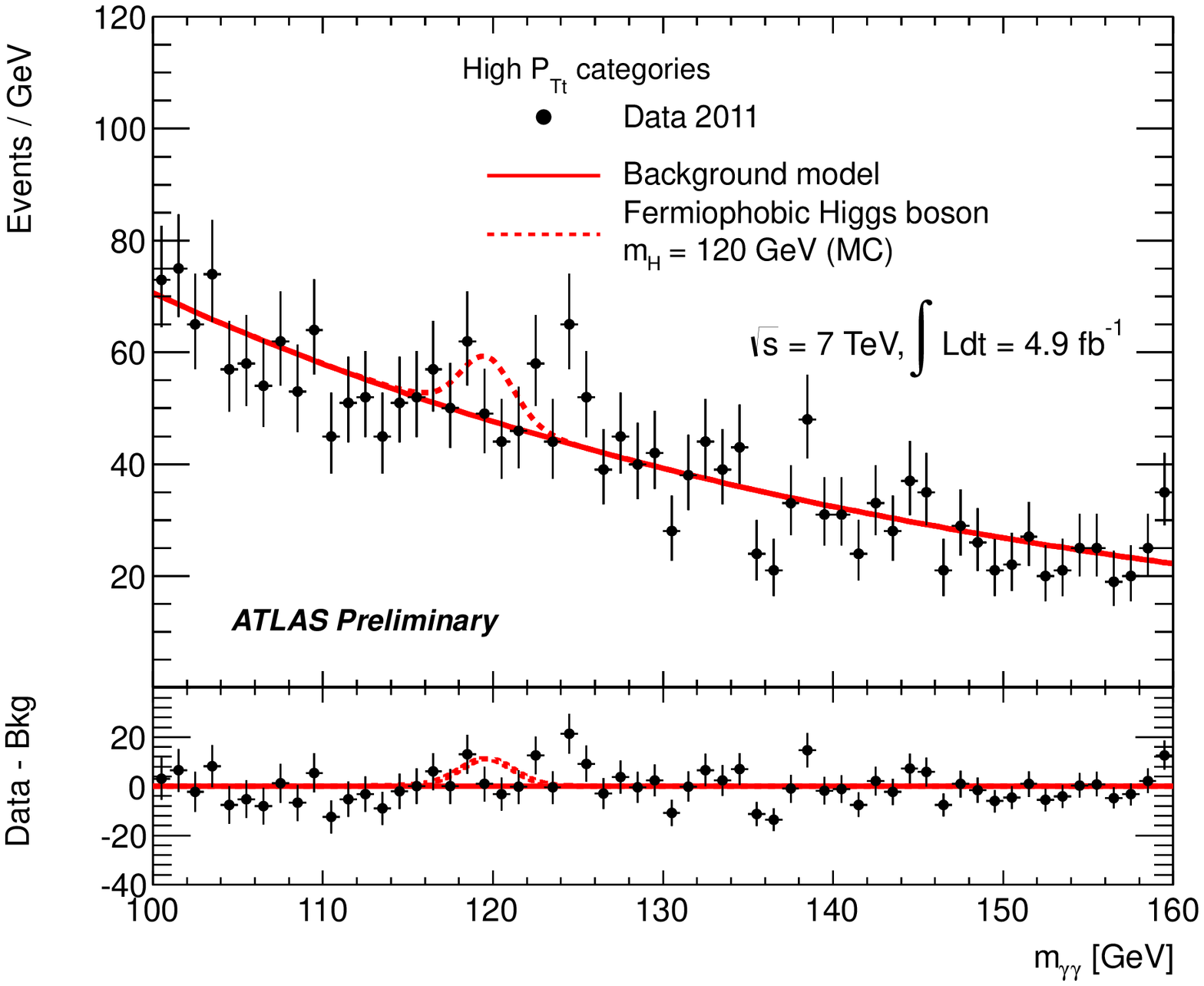,height=\fheighta}}
\hfill
\subfigure[]{\psfig{figure=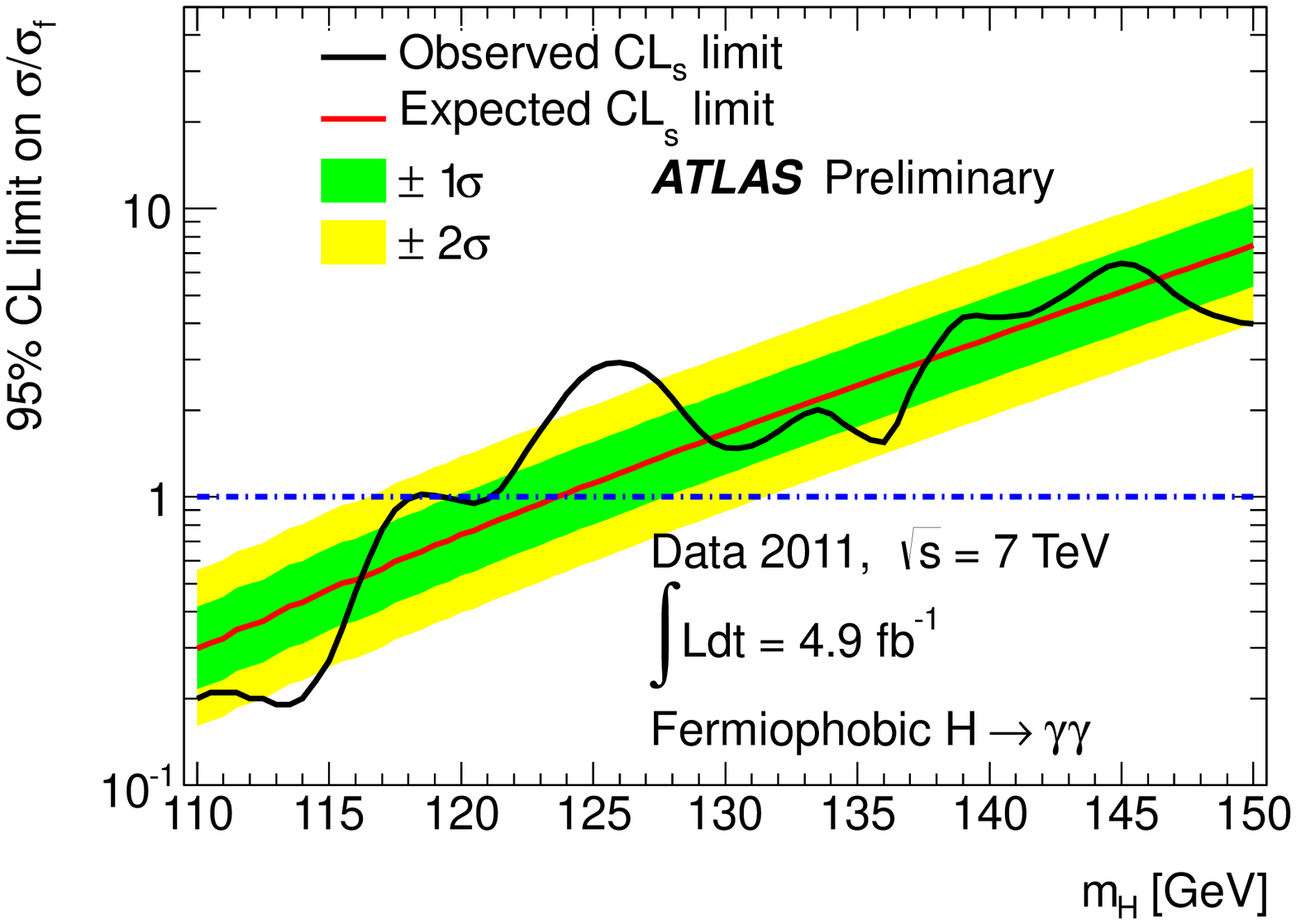,height=\fheighta}}
\hfill\ 
\caption{Fermiophobic $H\to\gamma\gamma$ search$^3$:
(a) diphoton invariant mass spectra for the high $p_{Tt}$  categories, overlaid with the background-only fit and signal expectation for $m_H=$120 GeV; 
(b) expected and observed 95\% confidence level limits normalized to the fermiophobic cross section times branching ratio expectation as a function of $m_H$.\hfill\
\label{fig:gg}}
\end{figure}

\begin{boldmath}
\section{Search for a fermiophobic Higgs boson via $H\to\gamma\gamma$}
\end{boldmath}
\noindent
The decay $H\to\gamma\gamma$ is searched for within a simple fermiophobic Higgs benchmark model in which the Higgs-fermion couplings are zero and SM couplings to bosons are assumed. The production cross section times the decay branching ratio for a fermiophobic $H\to\gamma\gamma$  is larger than in the SM for Higgs boson masses below 120 GeV.  The experimental sensitivity is further enhanced 
due to the fact  that a fermiophobic Higgs boson can only be produced through vector boson fusion and associated production with vector bosons, leading to typically larger transverse momenta of the Higgs boson and its decay products than for the dominant SM production via $gg$ fusion. The analysis~\cite{hgg} uses 4.9~fb$^{-1}$ of ATLAS data and follows the same procedure as the SM Higgs search in this decay channel~\cite{smhgg} . 
Pairs of isolated high-$p_T$ photons in the invariant mass range 100~GeV$<m_{\gamma\gamma}<$160~GeV are analyzed in nine categories according to the presence of photon conversions, the photon calorimeter impact point, and the diphoton transverse momentum $p_{Tt}$ orthogonal to the diphoton thrust axis in the transverse plane. 
 Signal events are expected to typically have a larger  $p_{Tt}$ than the background. Figure~\ref{fig:gg}(a) shows the $m_{\gamma\gamma}$ distribution for high-$p_{Tt}$ events compared to the background and signal expectations. The resulting cross section limits, see Fig.~\ref{fig:gg}(b),  exclude a fermiophobic Higgs boson in the mass ranges [110.0,118.0] and [119.5,121.0], with an expected exclusion range of  [110.0,123.5]. The largest excess over the background-only hypothesis is found at  a Higgs mass $m_H=125.5$~GeV, which however corresponds to only $1.6\sigma$ when the look-elsewhere effect is considered.  Since this conference, the corresponding final results of this analysis have been submitted for publication~\cite{hggpub}.\\

 
\begin{boldmath}
\section{Search for MSSM neutral Higgs bosons via $H/A/h\to\tau\tau$}
\end{boldmath}
\noindent
The minimal supersymmetric extension to the Standard Model (MSSM)  comprises two Higgs doublets of opposite weak hypercharge. This results in five observable Higgs bosons, three of which (the $CP$-even $h$, $H$, and the $CP$-odd $A$) are electrically neutral and two are charged ($H^\pm$). The decays to $\tau$ leptons provide particularly important search channels because the Higgs couplings to third-generation down-type fermions are strongly enhanced for large regions of the MSSM parameter space. 
\begin{figure}[b!]
 \hfill
\subfigure[]{\psfig{figure=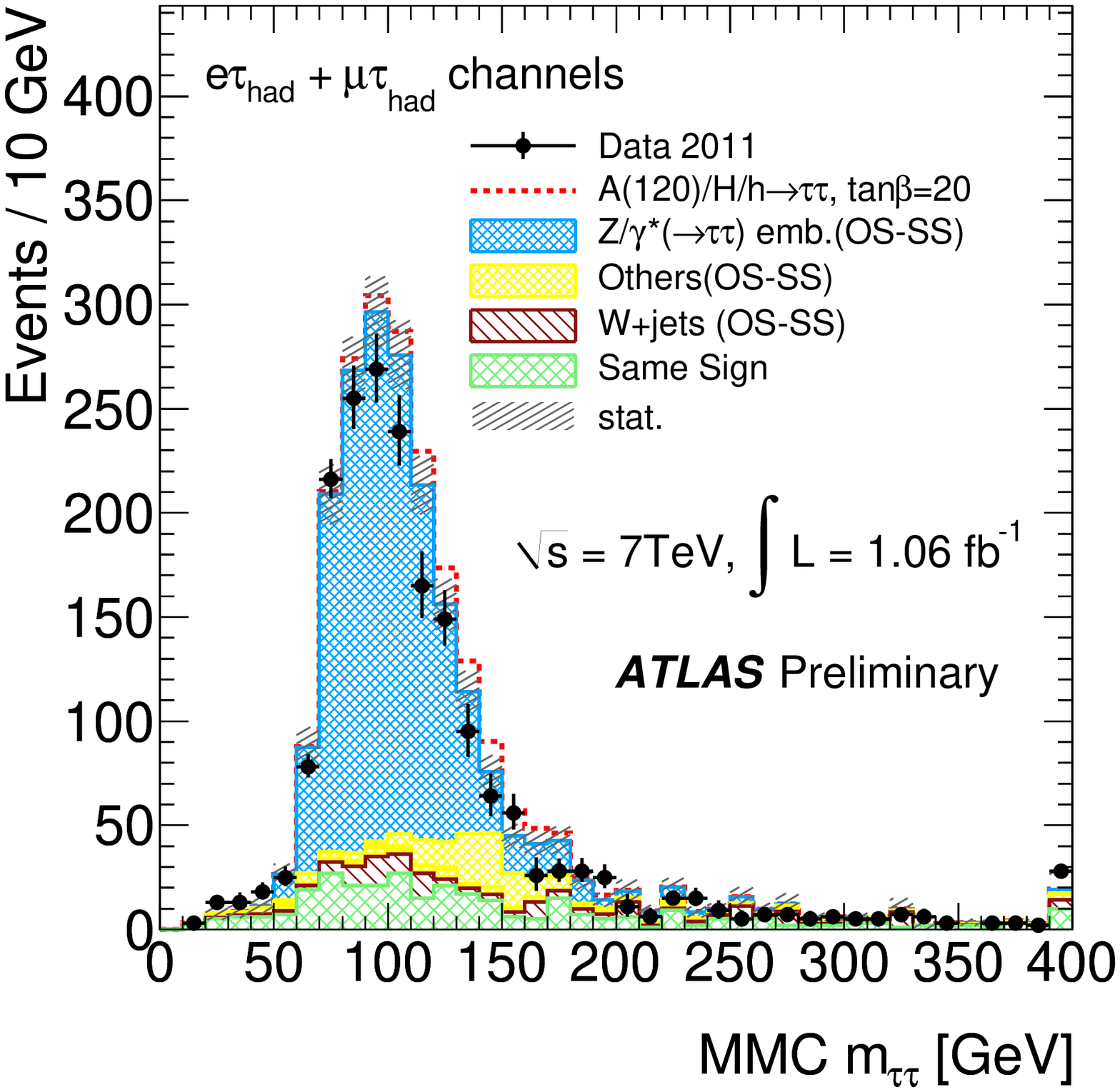,height=\fheightb}}
\hfill
\subfigure[]{\psfig{figure=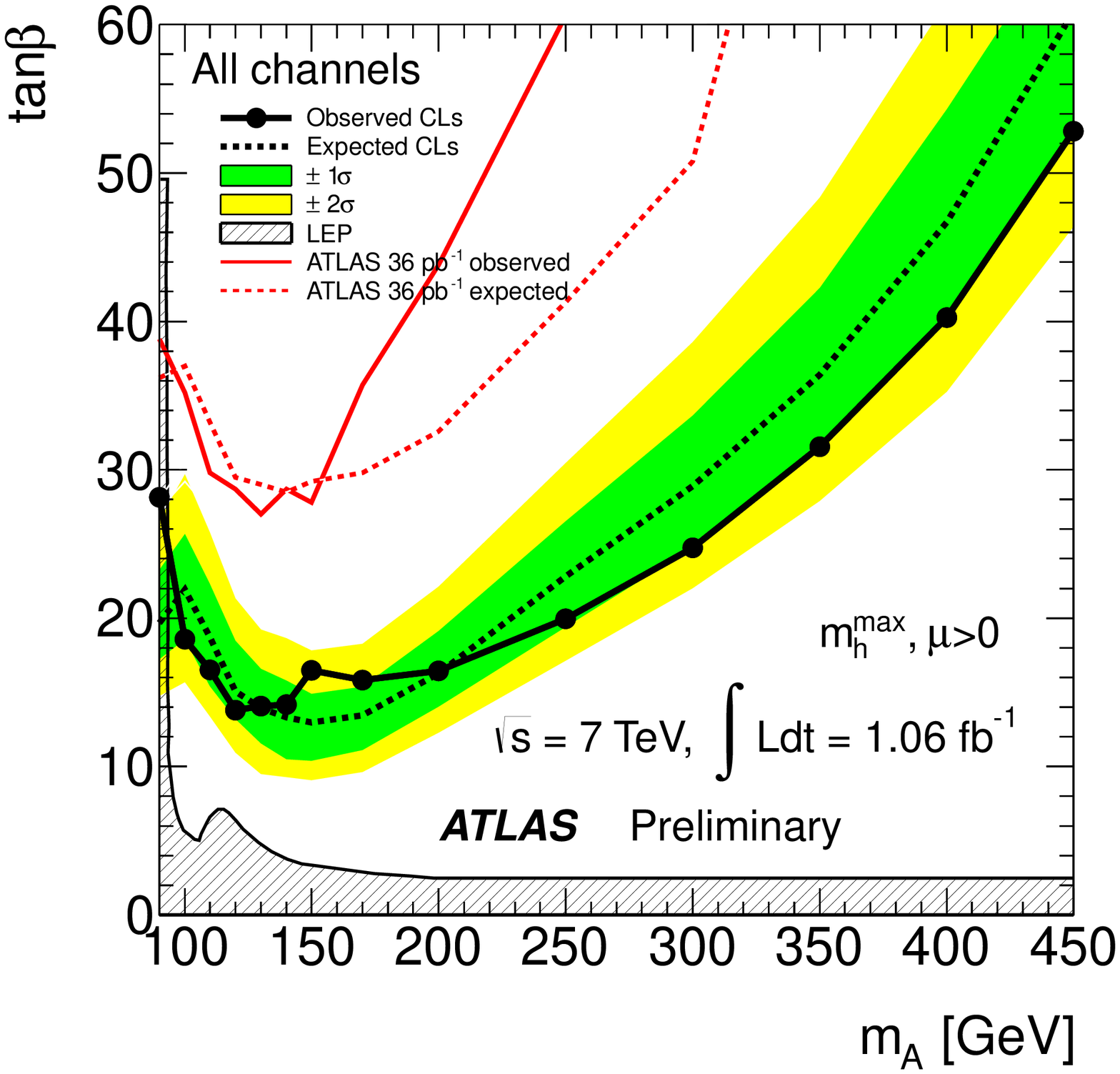,height=\fheightb}}
\hfill\
\caption{MSSM $Higgs\to\tau\tau$ search$^6$:
(a) $\tau\tau$ (MMC) mass distribution for the $\ell\tau_{had}$ final state. The data are compared with the background expectation and an added hypothetical MSSM signal ($m_A=120$~GeV, $\tan\beta=20$);
(b) expected and observed exclusion limits in the $m_A$-$\tan\beta$ plane of the MSSM derived from the combination of all channels.
The region above the limit curve is excluded at the 95\% confidence level. 
\hfill\
\label{fig:nh}}
\end{figure}
The most recent ATLAS results~\cite{htt} on the search for MSSM neutral Higgs bosons decaying into a pair of $\tau$ leptons are based on 2011 data corresponding to an integrated luminosity of $\mathcal{L}$=1.06~fb$^{-1}$. The analysis considers the final states $\tau\tau\to e\mu$, $\ell\tau_{had}$ ($\ell=e$ or $\mu$), and $\tau_{had}\tau_{had}$, where $\tau_{had}$ denotes a hadronically decaying $\tau$ lepton.  After signal selection, 4630 events are observed in this data sample. The observed number of events is consistent with the expected background of $4900\pm 600$ events. Corresponding exclusion limits are obtained from the $m_{\tau\tau}$ distribution, which for the dominant 
 $\ell\tau_{had}$  channel is reconstructed using the so-called missing-mass calculator (MMC) technique~\cite{mmc}, see Fig.~\ref{fig:nh}(a). Data control samples are used, where possible, to estimate or validate the background distributions; this is particularly relevant for the irreducible $Z\to\tau\tau$ background, which is modeled by embedding simulated $\tau$ decays in selected $Z\to\mu\mu$ data events.
 Fig.~\ref{fig:nh}(b) shows the resulting limits in the context of the MSSM $m^\mathrm{max}_h$ scenario~\cite{mhmax} as a function of the $CP$-odd Higgs boson mass $m_A$ and the ratio $\tan\beta$ of the vacuum expectation values of  the two Higgs doublets.\\

 
\begin{boldmath}
\section{Search for a MSSM charged Higgs boson via $H^\pm\to\tau\nu_\tau$}
\end{boldmath}
\noindent
The discovery of a charged Higgs boson $H^\pm$ would clearly establish physics beyond the SM. For charged Higgs 
\begin{figure}[b!]
\hfill
\subfigure[]{\psfig{figure=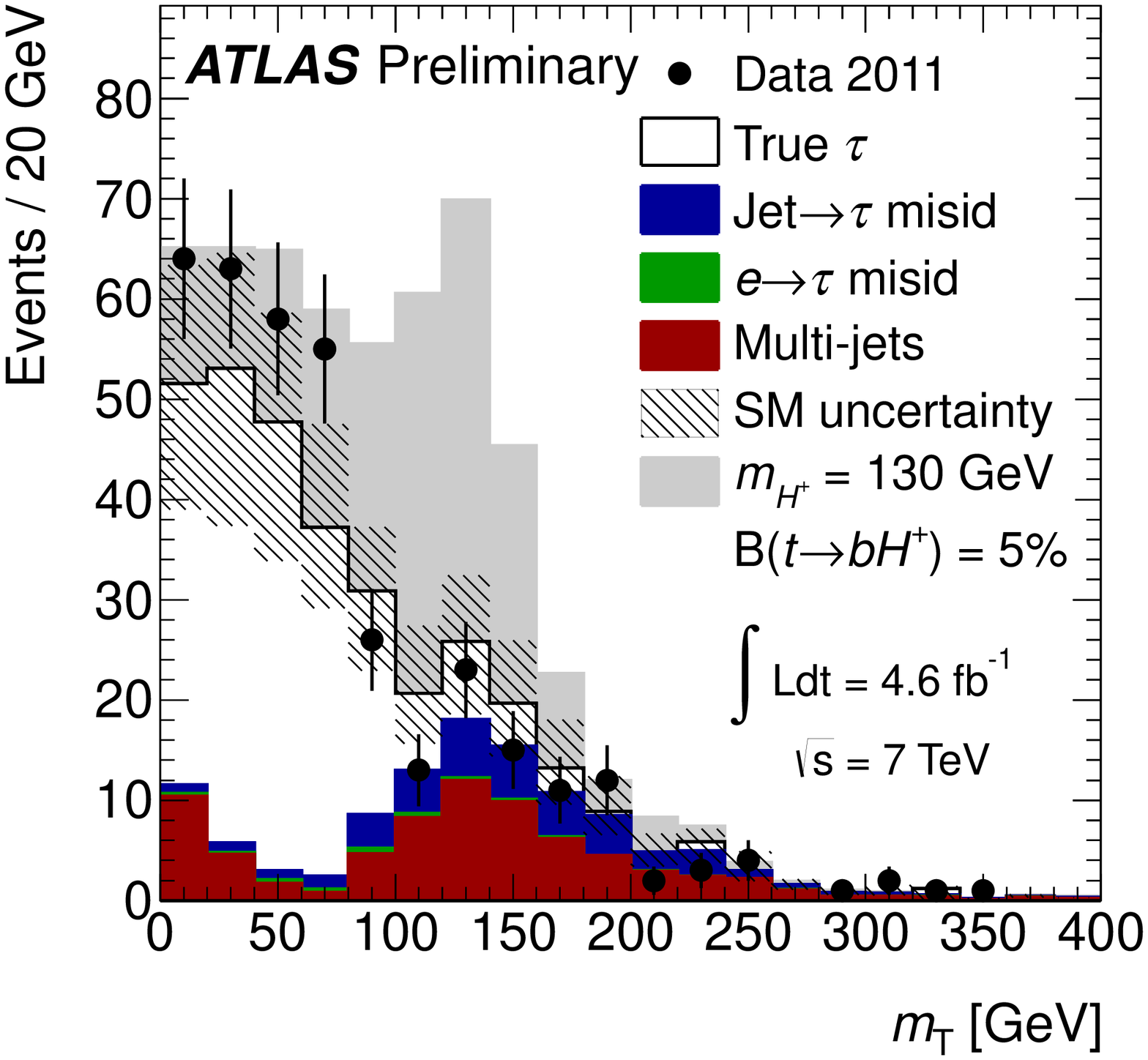,height=\fheightb}}
\hfill
\subfigure[]{\psfig{figure=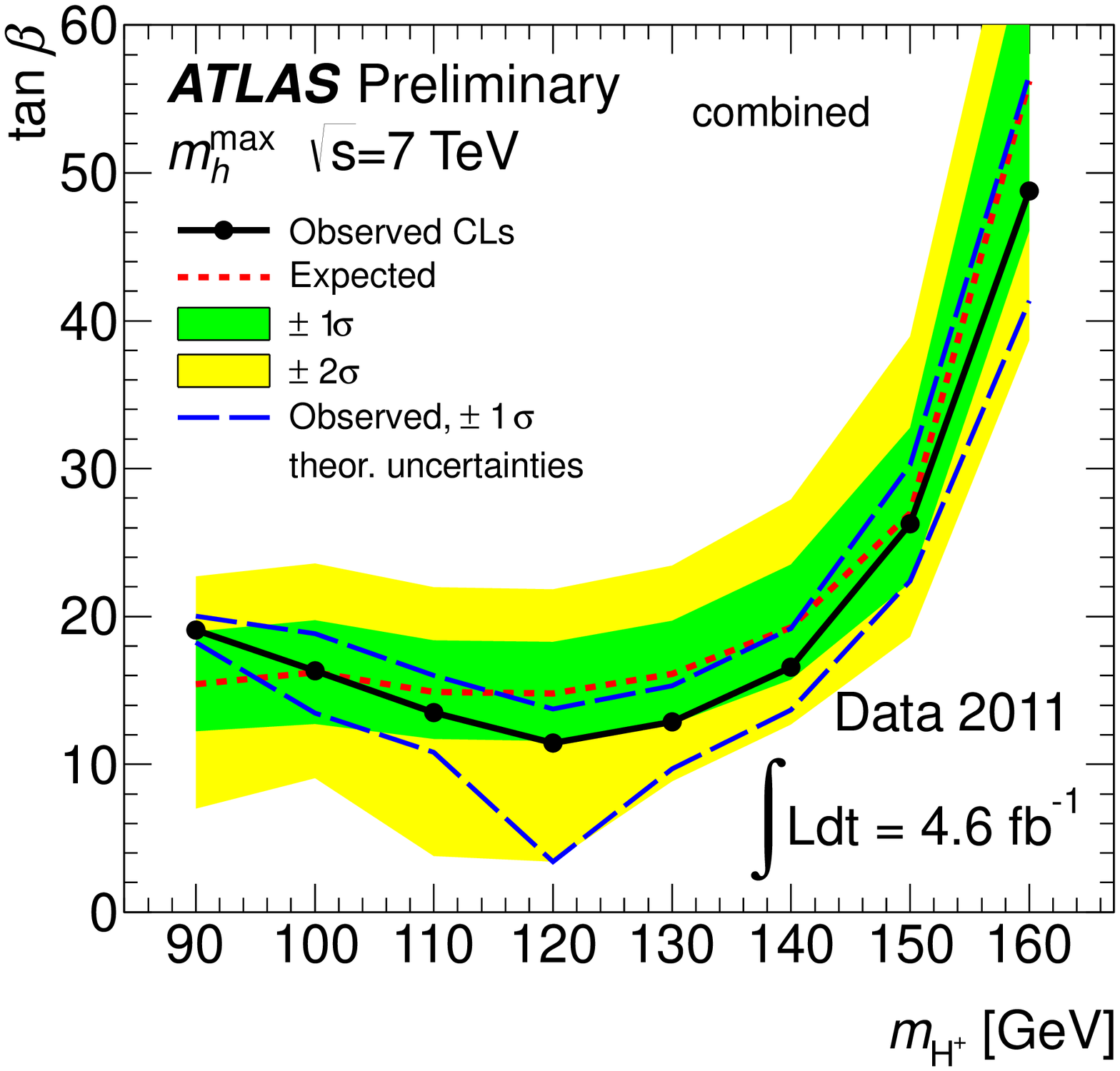,height=\fheightb}}
\hfill\
\caption{MSSM $H^\pm\to\tau\nu_\tau$ search$^9$:
(a) distribution of $m_T$ after all selection cuts in the $\tau$+jets channel. 
The stacked histogram shows the predicted contribution of signal+background for 
$m_{H^+} = 130~\mbox{GeV}$, assuming ${\cal B}(t \rightarrow bH^{\pm}) = 5\%$ and ${\cal B}(H^\pm \rightarrow \tau\nu_\tau) = 100\%$;  
(b) 95\% CL exclusion limits on $\tan\beta$ as a function of $m_{H^\pm}$. \hfill\
\label{fig:ch}}
\end{figure}
masses smaller than the top quark mass, the dominant production mechanism is the production of a $t\bar{t}$ pair with subsequent decay of one of the top quarks to a $b$ quark and a charged Higgs boson, which in turn predominantly decays via $H^\pm\to\tau\nu_\tau$ if $\tan\beta>3$. A recent ATLAS search~\cite{hpt}, based on the entire 2011 data set ($\mathcal{L}=4.6~$fb$^{-1}$), considers the decays 
$t\bar{t}\to b\bar{b}W^\mp H^\pm\to b\bar{b}(q\bar{q}')(\tau_{lep}\nu_\tau)$, 
$b\bar{b}(\ell\nu_\ell)(\tau_{had}\nu_\tau)$, and
$b\bar{b}(q\bar{q}'))(\tau_{had}\nu_\tau)$, 
which in the context of this analysis are referred to as lepton+jets, $\tau$+lepton and $\tau$+jets channels, respectively.
Different discriminating variables are used for the individual channels. For lepton+jets final states, the angular correlation between the $b$ jet and the charged lepton coming from the same top quark candidate is exploited; also, a transverse mass $m^H_T$ is reconstructed providing 
an event-by-event lower bound on the mass of the leptonically decaying charged ($W $or Higgs) boson produced in the top quark decay. In the $\tau$+lepton and $\tau$+jets channels, the distributions of the missing transverse energy ($E^\mathrm{miss}_T$) and the transverse mass $m_T$ of the $\tau$-$E^\mathrm{miss}_T$ system (cf.~Fig.~\ref{fig:ch}(a)), respectively, are used for the statistical analysis.
In all cases the data are found to be consistent with the expected SM background. Assuming 
that the branching fraction $B(H^\pm\to\tau\nu)$ is 100\%, this leads to upper limits on $B(t\to bH^\pm)$
between 5\% and 1\% for charged Higgs boson masses $m_{H^+}$ ranging from 90 to 160~GeV,
respectively. Within the $m^\mathrm{max}_h$  scenario of the MSSM, values of $\tan\beta$ larger than 
13-26 are excluded for charged Higgs boson masses in the range 90~GeV$<m_{H^\pm} <$150~GeV as shown in Fig.~\ref{fig:ch}(b).
Since this conference, the corresponding final results of this analysis have been submitted for publication~\cite{hptpub}.\\ 

\begin{figure}[h!]
\hfill
\subfigure[]{\psfig{figure=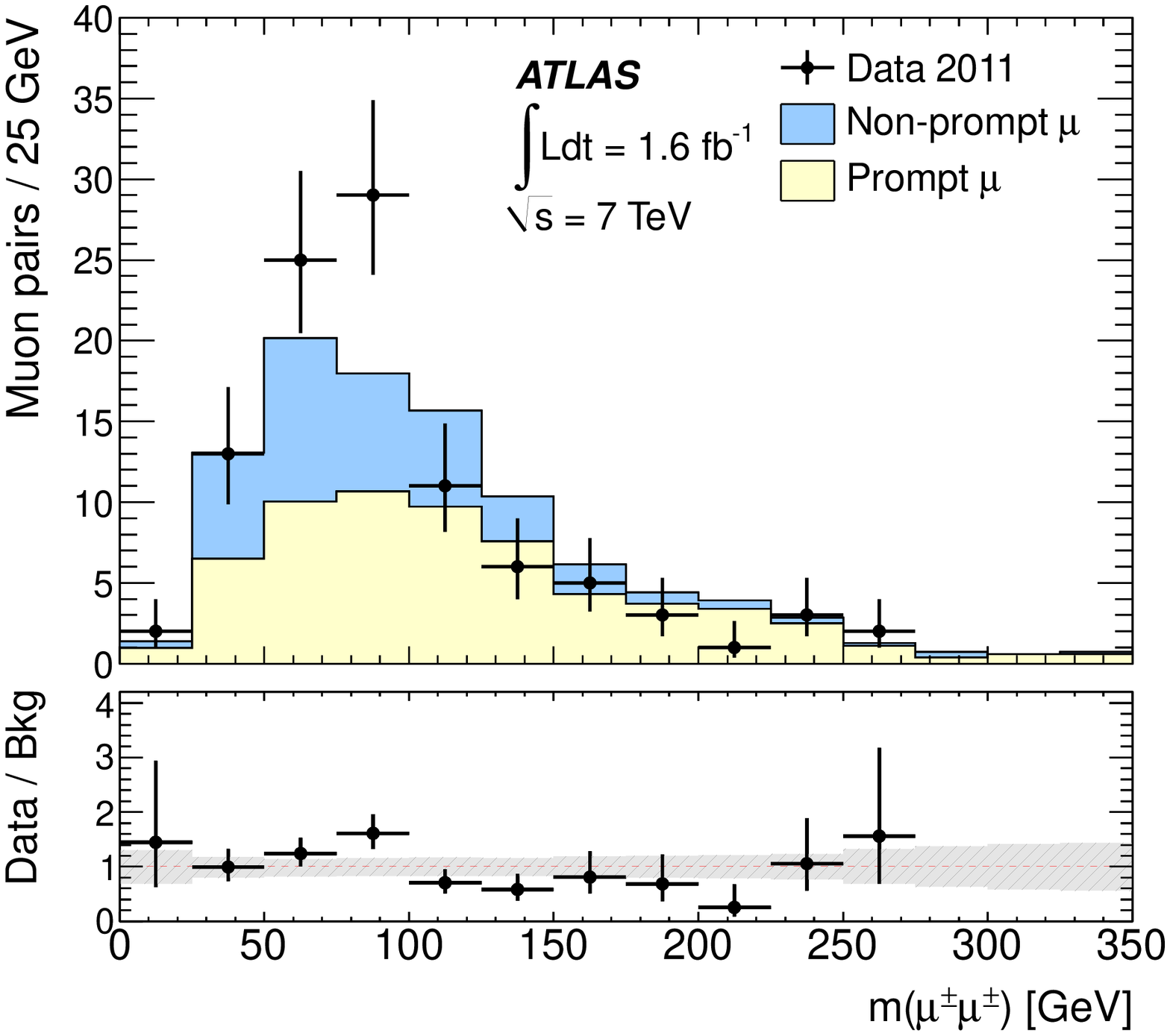,height=\fheighta}}
\hfill
\subfigure[]{\psfig{figure=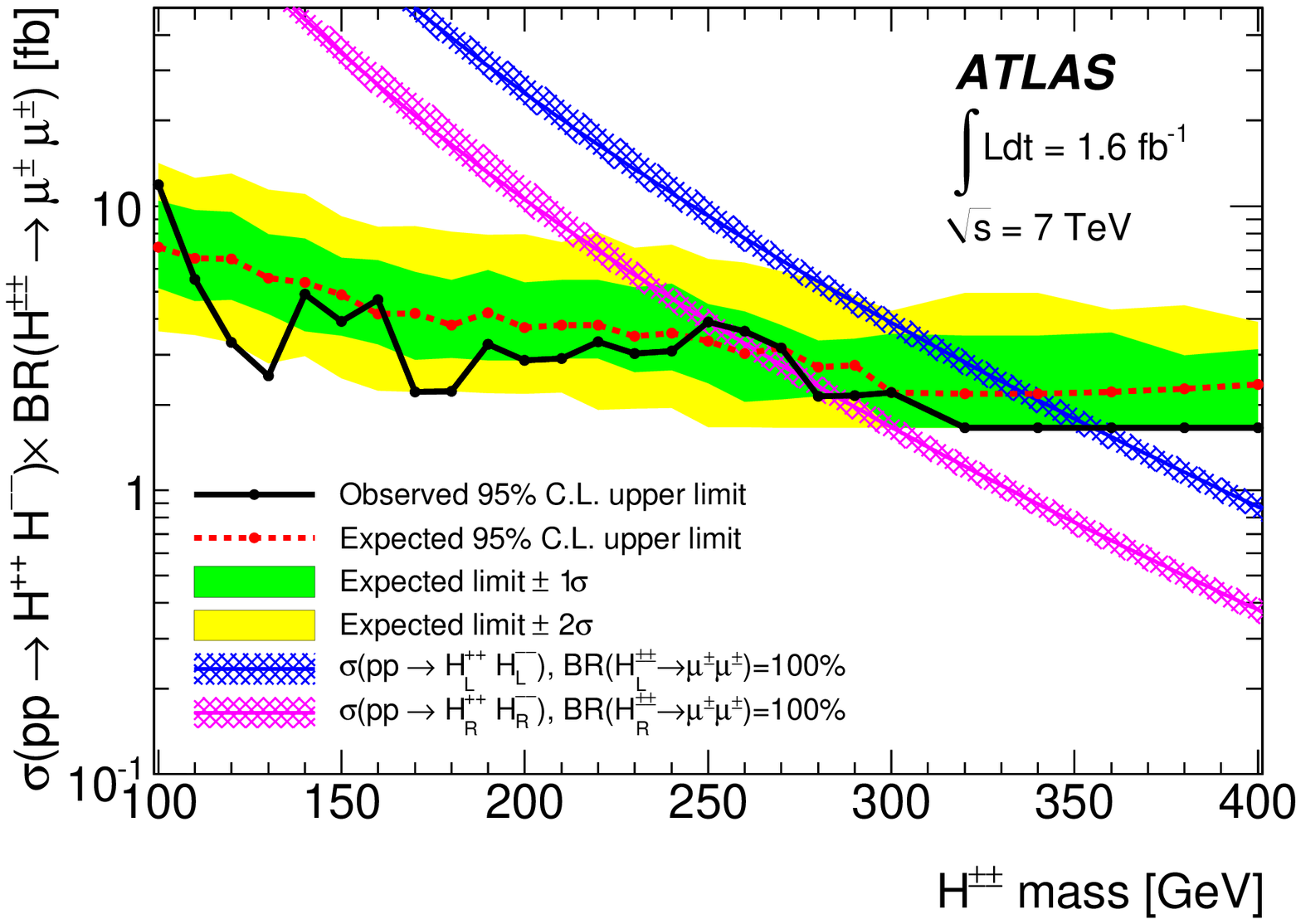,height=\fheighta}}
\hfill\ 
\caption{$H^{\pm\pm}\to\mu^{\pm}\mu^{\pm}$ search$^{11}$:
(a) distribution of the dimuon invariant mass for $\mu^\pm\mu^\pm$ pairs. The data are compared to the estimated background; 
(b) Upper limit at 95\% C.L. on the cross section times branching ratio for pair production of doubly charged Higgs bosons decaying to two muons. Superimposed is the predicted cross section for $H^{++}_RH^{--}_R$ and $H^{++}_LH^{--}_L$ production assuming a branching ratio to muons of 100\%. The bands on the predicted cross sections corresponds to the theoretical uncertainty of 10\%.\hfill\ 
\label{fig:dch}}
\end{figure}

\begin{boldmath}
\section{Search for a doubly-charged Higgs boson in same-sign dimuon final states}
\end{boldmath}
\noindent
Going beyond the MSSM, there are a number of scenarios, such as Higgs triplet and left-right-symmetric models, predicting doubly-charged Higgs bosons $H^{\pm\pm}$. ATLAS has performed a search~\cite{hpp} for the decay $H^{\pm\pm}\to\mu^{\pm}\mu^{\pm}$ using 2011 data corresponding to a luminosity of 1.6~fb$^{-1}$ in the context of an inclusive analysis of dimuon pairs with the same electric charge, where a doubly-charged Higgs boson could be observed as a narrow resonance in the dimuon mass spectrum. Figure~\ref{fig:dch}(a) shows the $m_{\mu\mu}$ distribution for selected pairs of same-sign muons with $p_T>20$~GeV and $|\eta|<2.5$; no significant deviation from the background expectation is observed. Assuming pair production of $H^{\pm\pm}$ bosons and a branching ratio to muons of 100\% (33\%), this analysis excludes masses below 355 (244) GeV and 251 (209) GeV for  $H^{\pm\pm}$ bosons coupling to left-handed and right-handed fermions, respectively, cf.~Fig.~\ref{fig:dch}(b). 


\end{document}